\documentclass[%
onecolumn,
10pt,
nofootinbib,
 amsmath,amssymb,
 aps,
 a4paper,
 notitlepage,
prd
]{revtex4-1}

\usepackage{graphicx}% Include figure files
\usepackage{dcolumn}% Align table columns on decimal point
\usepackage{bm}% bold math
\usepackage{color}
\usepackage[utf8]{inputenc} 

\begin{document}

\title{Stable Cosmology in Generalised Massive Gravity}% Force line breaks with \\

\author{Michael Kenna-Allison}
\email{michael.kenna-allison@port.ac.uk}
\affiliation{Institute of Cosmology and Gravitation, University of Portsmouth\\ Dennis Sciama
Building, Portsmouth PO1 3FX, United Kingdom}

\author{A. Emir G\"umr\"uk\c{c}\"uo\u{g}lu}
\email{emir.gumrukcuoglu@port.ac.uk}
\affiliation{Institute of Cosmology and Gravitation, University of Portsmouth\\ Dennis Sciama
Building, Portsmouth PO1 3FX, United Kingdom}

\author{Kazuya Koyama}
\email{kazuya.koyama@port.ac.uk}
\affiliation{Institute of Cosmology and Gravitation, University of Portsmouth\\ Dennis Sciama
Building, Portsmouth PO1 3FX, United Kingdom}

\date{\today}% It is always \today, today,

\begin{abstract}
We present a thorough analysis of the cosmological perturbations in Generalised Massive Gravity. This is an extension of de Rham-Gabadadze-Tolley theory where the translation invariance in the St\"uckelberg field space is broken. This allows the mass parameters to be promoted to functions of the St\"uckelberg fields. We consider an exact cosmological background in this theory and study the stability of perturbations. We derive conditions to avoid ghost, gradient and tachyonic instability. The cosmology is an extension of the self-accelerating branch of the constant mass parameter theory, but now all five massive graviton polarisations propagate.
For concreteness, we consider a minimal version of the theory where cosmology undergoes an accelerated expansion at late times and show that the perturbative stability is preserved for a range of parameters.
\end{abstract}
\maketitle

%\tableofcontents

\section{Introduction}

The studies of infra-red modifications to General Relativity (GR) on large distances has been partly fuelled by the discovery of the late-time acceleration of the universe \cite{Perlmutter:1998np,Riess:1998cb}. Giving a mass to graviton offers an arguably more natural explanation of the late-time acceleration of the universe than the cosmological constant. Historically, it has been a challenge to formulate an interacting massive gravity theory with $5$ degrees of freedom. The first attempt was the linear massive spin--2 theory by Fierz and Pauli  \cite{Fierz:1939ix}, and it was quickly realised that the theory needed a non-linear completion to be continuous with General Relativity (GR) \cite{V1,Zakharov:1970cc,Vainshtein}. However, generic non-linear extensions have an extra degree of freedom that causes the Boulware-Deser ghost pathology \cite{PhysRevD.6.3368}. 

A non-linear theory without the Boulware-Deser mode was constructed by de Rham, Gabadadze and Tolley (dRGT) \cite{PhysRevLett.106.231101,PhysRevD.82.044020} who proposed four independent potential terms specially tuned to recover a stable Minkowski background with $5$ propagating degrees of freedom, as expected from a massive spin--2 field. The theory is written in a manifestly covariant form, using four scalar fields $\phi^a$ that are responsible for breaking diffeomorphisms and generating the mass. The fields enjoy an internal Poincar\'e symmetry, which then ensures that in the unitary gauge one has Lorentz invariance. The symmetry also implies that the \emph{fiducial metric} $f_{\mu\nu} \equiv \eta_{ab} \partial_\mu \phi^a \partial_\nu \phi^b$ is the only space-time tensor (other than the physical metric) that appears in the action. The mass terms then are written as an interaction between the physical metric $g$ and the fiducial metric via the $g^{-1}f$ tensor. 

Despite the theoretical success of dRGT theory, exact Friedmann-Lema\^itre-Robertson-Walker (FLRW) backgrounds suffer from a variety of problems
\footnote{Approximately FLRW backgrounds may be safe against perturbations \cite{DAmico:2011eto,DeFelice:2013awa}, although the lack of overall homogeneity and/or isotropy poses a technical challenge for obtaining accurate predictions beyond the background level. There also exist exact FLRW solutions that have a spatially isotropic fiducial metric \cite{Koyama:2011yg, Gratia:2012wt} but these solutions are likely to be unstable  \cite{Motloch:2015gta}. It was argued in \cite{Hogas:2019ywm} that  linear perturbation theory is not sufficient to study cosmological perturbations.}. 
In general two branches are allowed: {\it i}. \emph{normal branch} which is a generalisation of the flat space-time solution, prohibits expansion \cite{DAmico:2011eto,Gumrukcuoglu:2011zh}; {\it ii}. \emph{self-accelerating branch}, where the mass term gives rise to an effective cosmological constant, suffers from a strong coupling problem \cite{Gumrukcuoglu:2011zh} and a non-linear ghost instability \cite{DeFelice:2012mx}. 

Giving up the Poincar\'e symmetry in the field space, one can generalise the theory without reviving the Boulware-Deser mode \cite{Hassan:2011tf}. However, for maximally symmetric field spaces, the cosmology is not free from issues. The fate of the self-accelerating branch is oblivious to the field space metric, while the normal branch does have different properties, although none positive, at least for maximally symmetric field spaces. For a massive gravity with a de Sitter field space, cosmology suffers from either a Higuchi ghost \cite{Higuchi:1986p1208} or fails to achieve a successful screening \cite{Fasiello:2012rw}. For anti de-Sitter, the cosmology cannot sustain an accelerated expansion \cite{MartinMoruno:2013gq}. 

The lack of stable cosmologies in dRGT massive gravity motivates extensions of the theory. One example is the bimetric theory \cite{Hassan:2011zd}, where the fiducial metric is promoted to a dynamical variable. Bimetric cosmology suffers from a gradient type instability \cite{Comelli:2012db,Konnig:2014dna,DeFelice:2014nja,Konnig:2014xva,Lagos:2014lca,Comelli:2015pua,Kenna-Allison:2018izo}, although resolutions do exist: a chameleon-like potential \cite{DeFelice:2017gzc}, screening as the instability takes over \cite{Aoki:2015xqa,Mortsell:2015exa}, hierachy between the two coupling constants \cite{Akrami:2015qga}, to name a few. Other extensions where an additional scalar field couples to the mass term provide a new way to achieve cosmology \cite{DAmico:2012hia,Huang:2012pe}, although ensuring perturbative stability continues to be a challenge \cite{Gumrukcuoglu:2013nza, DAmico:2013saf,Anselmi:2017hwr,Gumrukcuoglu:2017ioy}.

One intriguing possibility is to modify the theory without introducing new degrees of freedom. By breaking the translation invariance in the field space, one can promote the mass parameters to functions of $\eta_{ab}\phi^a\phi^b$, while preserving Lorentz invariance
\footnote{If one allows violations of Lorentz invariance, the number of degrees of freedom for the massive graviton can be as small as $2$ \cite{DeFelice:2015hla}.}.
 This construction preserves the dRGT tuning that removes the Boulware-Deser mode \cite{deRham:2014lqa}. FLRW solutions were reported to be stable in some decoupling limit and all 5 graviton polarisations were shown to propagate, evading the strong coupling problem of dRGT massive gravity\cite{deRham:2014gla}.

In this work, we perform a full stability analysis of cosmology in GMG with a k-essence matter as a model for an irrotational barotropic perfect fluid. We calculate the dispersion relations for the scalar, vector and tensor graviton modes, then identify the conditions for avoiding ghost, gradient and tachyonic instabilities. For concreteness, we fix the functions to allow a small deviation from the constant parameter dRGT theory and show that all stability conditions can be satisfied simultaneously for a self-accelerating background in this minimal version of the theory.

The paper is organised as follows. In Section \ref{model} we outline the theory and the field configuration. In Section \ref{BGcos} we describe the background dynamics and in Section \ref{pert} we describe the procedure for calculating the quadratic action and outline the stability conditions. Section \ref{mini} is where we introduce the minimal theory with small deviations from dRGT. We conclude with Section \ref{conc} where we summarise our results and discuss future work.

\section{The Set-Up}\label{model}
In this section we review the Generalised Massive Gravity theory and discuss the field configuration for cosmological solutions. 
The gravitational action consists of the Einstein-Hilbert term and the generalised mass terms
\begin{equation}\label{model1}
S=\frac{M_p^2}{2}\int d^4 x \sqrt{-g}\left[R+2m^2\sum_{n=0}^4\alpha_n(\phi^a\phi_a)\;\mathcal{U}_n\left[\mathcal{K}\right]\right]+ S_{matter}\,,
\end{equation}
where $\mathcal{U}_n$ are the dRGT potential terms,
\begin{align}
\mathcal{U}_0(\mathcal{K}) &= 1\,,\nonumber\\
\mathcal{U}_1(\mathcal{K}) &= [\mathcal{K}]\,,\nonumber\\
\mathcal{U}_2(\mathcal{K}) &= \frac{1}{2!}\,([\mathcal{K}]^2-[\mathcal{K}^2])\,,\nonumber\\
\mathcal{U}_3(\mathcal{K}) &= \frac{1}{3!}\,([\mathcal{K}]^3-3[\mathcal{K}][\mathcal{K}^2]+2[\mathcal{K}^3])\,,\nonumber\\
\mathcal{U}_4(\mathcal{K}) &= \frac{1}{4!}\,([\mathcal{K}]^4-6[\mathcal{K}]^2[\mathcal{K}^2]+8[\mathcal{K}][\mathcal{K}^3]+3[\mathcal{K}^2]^2-6[\mathcal{K}^4])\,.
\end{align}
Here, square brackets denote trace operation and the tensor $\mathcal{K}$ is defined by,
\begin{equation}
\mathcal{K}_{\nu}^{\mu}=\delta_{\nu}^{\mu}-\left(\sqrt{g^{-1}f}\right)_{\nu}^{\mu}\,.
\end{equation} 
This tensor refers to the space-time tensor $f_{\mu\nu}$, the fiducial metric, which is defined as
\begin{equation}
f_{\mu\nu} \equiv \eta_{ab}\partial_\mu \phi^a \partial_\nu\phi^b\,,
\end{equation}
with $(a,b = 0,1,2,3)$.
In standard dRGT massive gravity, $\phi^a$ are the St\"uckelberg fields, arising from the reintroduction of diffeomorphism invariance. However, if the translation invariance in the field space is broken, the four fields can also appear in the Lorentz invariant combination $\eta_{ab}\phi^a\phi^b$; in GMG theory, the mass parameters $\alpha_n$ are promoted to functions of this combination \cite{deRham:2014gla}.

For the matter sector we consider a k-essence field with action
\begin{equation}\label{matter}
S_{matter}=\int d^4 x \sqrt{-g}\;P(X),
\end{equation}
where
\begin{equation}
X=-g^{\mu \nu}\partial_{\mu}\varphi\partial_{\nu}\varphi\,.
\end{equation}

In order to achieve an isotropic and homogeneous universe for both the physical and fiducial metric, we need $f_{\mu\nu}$ to have the same FLRW symmetries as $g_{\mu\nu}$ in the same coordinate system, since they are coupled via $g^{-1}f$. Moreover, we also need to insure that $\phi^a\phi_a $ stays uniform. For a Minkowski field space metric, the unique field configuration that is compatible with these symmetries is 
\begin{align}\label{fields}
\phi^0&=f(t)\sqrt{1+\kappa(x^2+y^2+z^2)}\,,\nonumber\\
\phi^1&=f(t)\sqrt{\kappa}x \,,\nonumber\\ 
\phi^2&=f(t)\sqrt{\kappa}y \,,\nonumber\\
\phi^3&=f(t)\sqrt{\kappa}z \,,
\end{align}
where $\kappa = |K| = -K$ is the absolute value of the negative constant curvature of the spatial slice. With this definition, the fiducial metric has the same form as an open FLRW solution \cite{Gumrukcuoglu:2011ew}
\begin{equation}\label{f}
f_{\mu \nu}dx^{\mu}dx^{\nu}=-\dot{f}(t)^2dt^2+\kappa f(t)^2 \Omega_{ij}dx^i dx^j\,,
\end{equation}
where an overdot denotes time derivative and $\Omega_{ij}$ is the metric of the constant time hypersurfaces with constant negative curvature
\begin{equation}
\Omega_{ij}dx^i dx^j=dx^2+dy^2+dz^2-\frac{\kappa (xdx+ydy+zdz)^2}{1+\kappa(x^2+y^2+z^2)}\,.
\end{equation}
Our metric ansatz is then an open FLRW
\begin{align}\label{g}
g_{\mu \nu}dx^{\mu}dx^{\nu}=-dt^2+a(t)^2 \Omega_{ij}dx^i dx^j\,.
\end{align}

We consider a uniform matter field background, i.e. $\varphi = \varphi(t)$. In this case, the k-essence can be interpreted as an irrotational fluid with pressure $P(\dot{\varphi}^2)$, while the energy density $\rho(\dot{\varphi}^2)$ and sound speed $c_s(\dot{\varphi}^2)$ of the analogue fluid is given by
\begin{equation}
\rho=2P'(\dot{\varphi}^2)\dot{\varphi}^2-P(\dot{\varphi}^2), \qquad c_s^2=\frac{P'(\dot{\varphi}^2)}{2P''(\dot{\varphi}^2)\dot{\varphi}^2+P'(\dot{\varphi}^2)}\,
\end{equation}
where a prime denotes differentiation with respect to the argument.

\section{Background Dynamics}\label{BGcos}
We now can use the field configurations outlined in the previous section to determine the background dynamics of cosmology. The total action in the mini-superspace approximation is 
\begin{equation}\label{BGaction}
S=\frac{M_p^2V}{2}\int N\,dt \,a^3\left[
-\frac{6\,\kappa}{a^2} - \frac{6\,\dot{a}^2}{a^2\,N^2} +2m^2\left(\alpha_0U_0+\alpha_1U_1+\alpha_2U_2+\alpha_3U_3+\alpha_4U_4\right)+\frac{2}{M_p^2}P(\dot\varphi^2) 
\right],
\end{equation}
where $\alpha_n=\alpha_{n}[-f(t)^2]$ and 
\begin{align}
U_0&=1\,,\nonumber\\
U_1&=4-\frac{3\,\sqrt{\kappa}\,f}{a}-\frac{\dot{f}}{N}\,,\nonumber\\
U_2&=3\,\left(1-\frac{\sqrt{\kappa}\,f}{a}\right)\left(2-\frac{\sqrt{\kappa}\,f}{a}-\frac{\dot{f}}{N}\right)\,,\nonumber\\
U_3&=\left(1-\frac{\sqrt{\kappa}\,f}{a}\right)^2\left(4-\frac{\sqrt{\kappa}\,f}{a}-\frac{3\,\dot{f}}{N}\right)\,,\nonumber\\
U_4&=\left(1-\frac{\sqrt{\kappa}\,f}{a}\right)^3\left(1-\frac{\dot{f}}{N}\right)\,.
\end{align}
Varying the action \eqref{BGaction} with respect to $N$, $a$ and $\varphi$, then fixing the cosmological time $N=1$, we get the following equations of motion \cite{deRham:2014gla}
\begin{align}\label{BGeqns}
3\,\left(H^2-\frac{\kappa}{a^2}\right)&=m^2L+\frac{\rho}{M_p^2}\,,\nonumber\\
2\left(\dot{H}+\frac{\kappa}{a^2}\right)&=m^2J(r-1)\xi-\frac{\rho+P}{M_p^2} \,,\nonumber\\
\dot{\rho}&=-3\,H\,(\rho+P)\,,
\end{align}
where for convenience, we defined
\begin{equation}
H \equiv \frac{\dot{a}}{a} \,,\qquad 
\xi \equiv\frac{\sqrt{\kappa}f}{a}\,,\qquad 
r\equiv \frac{a\,\dot{f}}{\sqrt{\kappa}f}\,.
\end{equation}
We also defined two combinations of the mass parameters
\begin{align}
\label{defLJ}
L & \equiv -\alpha_0+(3\,\xi-4)\alpha_1-3\,(\xi-1)(\xi-2)\alpha_2+(\xi-1)^2(\xi-4)\alpha_3+(\xi-1)^3\alpha_4
\,,\nonumber\\
J&\equiv\alpha_1+(3-2\xi)\alpha_2+(\xi-1)(\xi-3)\alpha_3+(\xi-1)^2\alpha_4\,,
\end{align}
where from the background equations, we infer that $m^2M_p^2L$ is the effective energy density arising from the mass term, while $m^2M_p^2J(1-r)\,\xi$ corresponds to the sum of the effective density and pressure. In standard dRGT, the quantity $J$ is forced to vanish, yielding a constant $\xi$ solution. As a result, the contribution to the Friedmann equation $m^2L$ becomes an effective cosmological constant. In contrast, in the GMG theory, this is no longer the case. By varying the action \eqref{BGaction} with respect to $f$, or equivalently using the contracted Bianchi identities, we obtain the St\"uckelberg constraint equation
\begin{equation}\label{stuck}
3\,H\,J\,(r-1)\xi-\dot{L}=0\,.
\end{equation}
Using the definition of $L$ from Eq.~\eqref{defLJ}, we can also rewrite this equation in the following form
\begin{equation}\label{stuck-alt}
3\,\left(H-\frac{\sqrt{\kappa}}{a}\right)\,J=-\frac{2\,a\,\xi}{\sqrt{\kappa}}
\left[
-\alpha'_0+(3\,\xi-4)\alpha'_1-3\,(\xi-1)(\xi-2)\alpha'_2+(\xi-1)^2(\xi-4)\alpha_3'+(\xi-1)^3\alpha_4'\right]\,.
\end{equation}
In this form, the dRGT limit can be trivially taken by $\alpha_n' \equiv \partial\alpha_n/\partial (-f^2) \to 0$. In this constant mass limit, the right hand side of \eqref{stuck-alt} vanishes, defining two branches of solutions: the normal branch with $H =\sqrt{\kappa}{a}$ which prevents expansion, while $J=0$ branch gives rise to a self-acceleration. The generalised mass term thus prevents the branching by breaking the factorised form of the constraint equation. Although this means that generically the mass term is no longer an effective cosmological constant, this also solves the problem of infinite strong coupling of scalar and vector perturbations, whose kinetic terms are proportional to $J$ in standard dRGT \cite{Gumrukcuoglu:2011zh}.

\section{Cosmological Perturbations}\label{pert}

To calculate the quadratic action we need to first introduce perturbations to the physical metric
\begin{equation}\label{decomp1}
g_{\mu \nu}dx^{\mu}dx^{\nu}=-(1+2\phi)dt^2+a\,(\partial_i B+B_i) \,dt\,dx^i+a^2\left(\Omega_{ij}+h_{ij}\right)dx^i dx^j,
\end{equation}
with $h_{ij}$ decomposed as
\begin{equation}\label{decomp}
h_{ij}=2\,\psi\, \Omega_{ij}+\left(D_i \,D_j-\frac{1}{3}\Omega_{ij}D_lD^l\right)E+\frac{1}{2}\left(D_i E_j+ D_j E_i\right)+\gamma_{ij}\,.
\end{equation}
Here, $D_i$ is the covariant derivative associated with the 3-metric $\Omega_{ij}$ and the spatial indicies are raised by the inverse metric $\Omega^{ij}$. The vectors in the above decomposition are divergence-free $D^i E_i = D^i B_i=0$, while the tensor is divergence and trace-free $D^i \gamma_{ij} = \Omega^{ij}\gamma_{ij} =0$. We also introduce matter perturbations through
\begin{equation}
\varphi = \varphi_0 +\delta\varphi\,,
\end{equation}
with quantities in the fluid analogue $P$, $\rho$ and $c_s$ all defined with respect to the background field. 
For the four scalar fields $\phi^a$ we exploit the diffeomorphism invariance to fix their perturbations to zero, depleting all the gauge freedom in the system.

In this decomposition, the scalars ($\phi$, $B$, $\psi$, $E$, $\delta\varphi$), vectors ($E_i$, $B_i$) and tensor ($\gamma_{ij}$) perturbations decouple at quadratic order in the action. We will therefore study them separately in the following.

\subsection{Tensors}

Starting with the tensor sector, we expand \eqref{model1} to second order in tensor perturbations, we find
\begin{equation}
S^{(2)}_T = \frac{M_p^2}{8}\int d^4x \,\sqrt{\Omega}\,a^3\,\left[\dot{\gamma}_{ij}\dot{\gamma}^{ij} + \frac{1}{a^2}\,\gamma_{ij}D_lD^l\gamma^{ij}+ \left(\frac{2\,\kappa}{a^2} -m^2\,\Gamma\right)\,\gamma_{ij}\gamma^{ij}\right]\,.
\end{equation}
We then expand $\gamma_{ij}$ in terms of tensor harmonics (see e.g. \cite{Kodama:1985bj})
\begin{equation}
\gamma_{ij} = \int k^2 dk\,\gamma_{|\vec{k}|} Y_{ij}(\vec{k},\vec{x})\,,
\end{equation}
with $D_lD^l Y_{ij} = -k^2 Y_{ij}$ and $D^iY_{ij} = \Omega^{ij}Y_{ij}=0$. This allows us to reduce the above action to
\begin{equation}
\label{tensoraction}
S^{(2)}_T = \frac{M_p^2}{8}\int dt d^3k \,\,a^3\,\Big[\vert\dot{\gamma}\vert^2  -\omega_T^2\,\vert\gamma\vert^2\Big]\,,
\end{equation}
where the tensor dispersion relation is
\begin{equation}
\omega_T^2=\left(\frac{k^2}{a^2}-\frac{2\kappa}{a^2}+m^2\;\Gamma\right)\,.
\end{equation}
The mass of the tensor mode $m \sqrt{\Gamma}$ is given in terms of the mass functions as
\begin{equation}
\Gamma\equiv \xi \left[\alpha_1+(3-2\xi)\alpha_2+(\xi-1)(\xi-3)\alpha_3+(\xi-1)^2\alpha_4\right] + (r-1)\,\xi^2\left[
-\alpha_2+(\xi-2)\alpha_3+(\xi-1)\alpha_4
\right]\,.
\end{equation}
This expression agrees with the standard dRGT tensor mass with constant $\alpha_n$ \cite{Gumrukcuoglu:2011zh}.

From \eqref{tensoraction} it is clear the tensors show no ghost or gradient instabilities since their kinetic term directly follows from a standard Einstein-Hilbert action. However one can place restrictions on $\Gamma$ requiring that $\Gamma>0$ to avoid a tachyonic instability. On the other hand, for $|m^2 \Gamma| \sim \mathcal{O}(H_0^2)$, the instability generically takes the age of the universe to develop, thus an imaginary mass is not necessarily a cause for concern.

\subsection{Vectors}

We next calculate the action \eqref{model1} at quadratic order in vector modes. The shift vector $B_i$ is non-dynamical so it can be integrated out by solving its algebraic equation of motion
\begin{equation}\label{bsol}
B_V=\frac{a(1+r)(k^2+2\kappa)}{2\left[(k^2+2\,\kappa)(r+1)+2m^2a^2J\,\xi\right]}\dot{E}_V\,,
\end{equation}
where we expanded the perturbations in terms of vector harmonics
\begin{equation}
B_i = \int k^2dk B_{V, |\vec{k}|} Y_i(\vec{k},\vec{x})\,,
\end{equation}
and similarly for $E_i$. Vector harmonics satisfy $D_iD^iY_j =-k^2\,Y_j$ and $D^iY_i =0$.
Upon substituting (\ref{bsol}) into the action, only one propagating vector remains,
\begin{equation}
\label{vectorquad}
S^{(2)}_V=\frac{M_p^2}{8}\int d^3k\; dt\; a^3\left[\mathcal{T} |\dot{E}_V|^2- \frac{k^2+2\kappa}{2}m^2\; \Gamma\; |E_V|^2 \right]\,,
\end{equation}
where the kinetic term is 
\begin{equation}
\mathcal{T}=\left(\frac{2}{k^2+2\,\kappa}+\frac{1+r}{m^2a^2J\,\xi}\right)^{-1}.
\end{equation}
In order to avoid ghost instability, the following condition must hold at sub-horizon scales
\begin{equation}
\mathcal{T}\Big\vert_{k\gg a\,H} = \frac{m^2\,a^2\,J\,\xi}{1+r}>0\,.
\end{equation}
The sound speed for the vector modes can be calculated by taking the ratio of the two terms in Eq.\eqref{vectorquad} in the sub-horizon limit
\begin{equation}
c_V^2 = \frac{m^2a^2\Gamma}{2 \,\mathcal{T}}\Big\vert_{k\gg a\,H} = \frac{(1+r)\Gamma}{2\,J\,\xi}.
\end{equation}
The squared-sound speeds should be positive to avoid gradient instability $c_V^2 \geq 0$.

\subsection{Scalars}
We now expand  the action \eqref{model1} up to quadratic order in scalar perturbations. Unfortunately, the full calculation involves expressions not suitable for presentation. However, we describe the procedure here and show some intermediate results.

At this stage we have an action with five degrees of freedom, $(\phi,\psi,B,E,\delta \varphi)$. However, the lapse and shift perturbations appear in the action without any time derivatives and can be integrated out. We first expand all perturbations in terms of scalar harmonics
\begin{equation}
\phi = \int k^2dk \phi_{S, |\vec{k}|} Y(\vec{k},\vec{x})\,,
\end{equation}
and similarly for other scalar perturbations. The scalar harmonics satisfy $D_iD^i Y = -k^2Y$. We solve the equation of motion for the shift $B$,
\begin{equation}\label{bsolve}
B=\frac{a\,(1+r)\left[3\,\delta \varphi(\rho+P)+M_p^2\dot{\varphi}\left[(k^2+3\kappa)\dot{E}+6(\dot{\psi}-H\phi)\right]\right]}{
3\,M_p^2\left[2\,(r+1)\kappa+m^2a^2J\,\xi\right]\dot{\varphi}}.
\end{equation}
where we omitted the subscript $S, |\vec{k}|$ and we will do so for all other perturbations in the following. 
Upon substituting the solution for $B$ back in the action, we then solve for the lapse perturbation $\phi$, which is,
\begin{align}\label{phisolve}
\phi=&\frac{M_p^2c_s^2\left[2\kappa(r+1)+m^2a^2J\xi\right]}{2\,M_p^2 c_s^2 H^2 \left[
2\,(k^2+3\,\kappa)(r+1)+3\,m^2\,a^2\,J\,\xi\right]
-\left[2\,\kappa\,(r+1)+m^2a^2J\,\xi\right](\rho+P)}\nonumber\\ 
&\times \Bigg\{ 
\frac{2\,k^2\,H\,(1+r)}{3\left[2\,\kappa\,(1+r)+m^2a^2J\,\xi\right]}\left[\frac{3\,(\rho+P)}{M_p^2\dot\varphi}\,\delta\varphi +(k^2+3\,\kappa)\dot{E} + \frac{3}{k^2}\left(2\,(k^2+3\,\kappa)+\frac{3\,m^2a^2J\,\xi}{1+r}\right)\dot{\psi}\right]\nonumber\\
&\qquad
+\frac{k^2(k^2+3\,\kappa)}{3\,a^2}E +\frac{2\,(k^2+3\,\kappa)+3\,m^2a^2J\,\xi}{a^2}\,\psi-\frac{(\rho+P)}{M_p^2c_s^2\dot{\varphi}}\,\delta\dot\varphi
\Bigg\}.
\end{align}

With the solution \eqref{phisolve} we reduce the quadratic action to a system with 3 degrees of freedom $(\psi,E,\delta\varphi)$. Formally, the action is the following;
\begin{equation}
S^{(2)}_S= \frac{M_p^2}{2}\int d^3 k\; dt\; a^3\left(\dot{\tilde{\Psi}}^\dagger\,\mathcal{\tilde{K}}\,\dot{\tilde{\Psi}}+\frac12\,\dot{\tilde{\Psi}}^\dagger\,\mathcal{\tilde{G}}\,\tilde{\Psi}+\frac12\,\tilde{\Psi}^\dagger\,\mathcal{\tilde{G}}^T\,\dot{\tilde{\Psi}}-\tilde{\Psi}^\dagger\,\mathcal{\tilde{M}}\,\tilde{\Psi}\right),
\end{equation}
where $\tilde{\Psi}\equiv(\psi,E,\delta\varphi)$ is the field array and where $\tilde{\mathcal{K}}$, $\tilde{\mathcal{G}}$ and $\tilde{\mathcal{M}}$ are the real $3\times 3$ kinetic, mixing and mass matrices respectively. The absence of the Boulware-Deser ghost implies that we can integrate out one more non-dynamical degree of freedom. Indeed, at this stage $\det \tilde{\mathcal{K}}=0$, indicating that there is at least one combination of the fields with vanishing kinetic term. To explicitly see this we define the quantity $Q$
\begin{equation}
Q\equiv \psi+\frac{k^2(r+1)(k^2+3\kappa)}{9m^2a^2J\,\xi+6(r+1)(k^2+3\kappa)}\,E-\left(\frac{H}{\dot{\varphi}}\right)\delta
\varphi\,.
\end{equation}
When we remove $\delta\varphi$ in favour of $Q$, the kinetic part of the action becomes diagonal and the non-dynamical nature of $\psi$ becomes manifest. 
We identify $\psi$ in this basis as the would-be Boulware-Deser mode and integrate it out. Unfortunately, the solution is not suitable to be presented here, but upon substitution into the action we obtain a system with two dynamical fields in a basis $\Psi=(Q,E)$, with the action formally
\begin{equation}
S^{(2)}_S= \frac{M_p^2}{2}\int d^3 k\; dt \left(\dot{\Psi}^\dagger\,\mathcal{K}\,\dot{\Psi}+\frac12\,\dot{\Psi}^\dagger\,\mathcal{G}\,\Psi + 
\frac12\,\Psi^\dagger\,\mathcal{G}^T\,\dot{\Psi}-\Psi^\dagger\,\mathcal{M}\,\Psi\right),
\label{finalscalar}
\end{equation} 
where $\mathcal{K}$, $\mathcal{G}$ and $\mathcal{M}$ are now the $2\times 2$ kinetic, mixing and mass matrices respectively in the new basis, with $\mathcal{K} = \mathcal{K}^T$, $\mathcal{M} = \mathcal{M}^T$.

\subsubsection{No Ghost Conditions}
The conditions for the absence of ghosts can be obtained by studying the positivity of the eigenvalues of the kinetic matrix $\mathcal{K}$ in the sub-horizon limit, which corresponds to taking $k \to \infty$. The two eigenvalues are determined as
\begin{equation}
e_1 = \mathcal{K}_{11}\,,\qquad
e_2=\frac{\det\;\mathcal{K}}{\mathcal{K}_{11}}\,.
\end{equation}
The exact expressions for the kinetic matrix eigenvalues are given in Appendix \ref{append}. In the dRGT limit $J, \dot{J} \to 0$, the second eigenvalue vanishes, in agreement with \cite{Gumrukcuoglu:2011zh}. However, with the varying mass functions, the strong coupling problem is resolved. 

We now expand the eigenvalues in the sub-horizon limit, obtaining
\begin{align}
e_1&=\frac{2\,(\rho+P)}{M_p^4c_s^2H^2} + \mathcal{O}\left(k^{-2}\right)\,,\nonumber\\
e_2&=\frac{3\,m^2a^4H}{2\,M_p^2r^2}\,\left[
\frac{r\,J\,\xi}{2\,H}\left(\frac{2\,\kappa}{a^2}-\frac{2\,\sqrt{\kappa}\,H}{a} -\frac{4\,H^2}{r}+m^2J\,\xi+\frac{\rho+P}{M_p^2}\right)+2\,H\,\Gamma-\dot{J}\,\xi
\right] + \mathcal{O}\left(k^{-2}\right)\,.
\end{align}
To avoid ghost instabilities, we need $e_1 >0$ and $e_2>0$. The first no-ghost condition is simply the null-energy condition, so we identify the first eigenmode as the matter sector. The second one is therefore the scalar graviton mode. However, from the subhorizon expression for $e_2$ we see that it no longer vanishes in the dRGT limit $J\,,\dot{J}\to 0$. The reason for this apparent discrepancy is that the sub-horizon limit does not commute with the dRGT limit. 

\begin{figure}[ht]
    \centering
    \includegraphics[width=0.5\columnwidth]{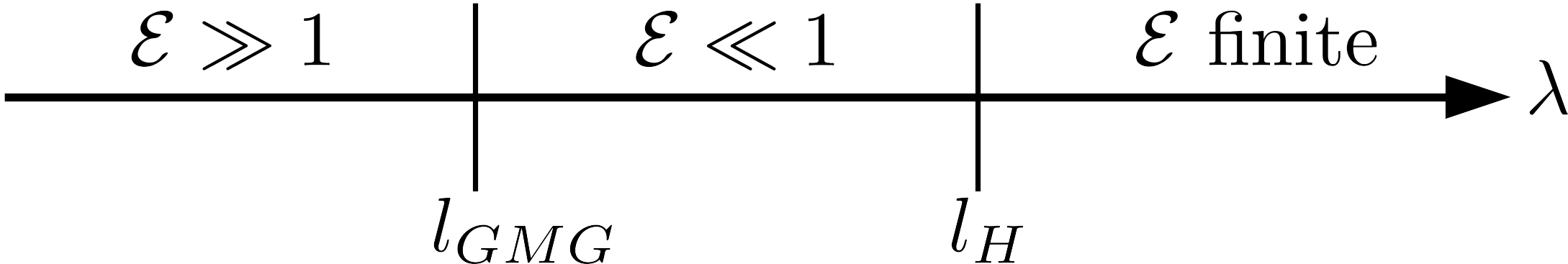}
    \caption{Schematic representation of different length scales. In this diagram, we assumed that $l_{GMG} < l_H$, which corresponds to small departures from standard dRGT.}
    \label{fig:scales}
\end{figure}

This peculiar behaviour can be understood by inspecting the terms in $e_2$. In the second of Eq.\eqref{eq:kinetic-exact}, there are some terms that vanish in any order of the limits. Neglecting these, we are left with two terms that dominate according to which limit is applied first
\begin{equation}
e_2\simeq \frac{m^2}{M_p^2}\left(
\frac{2(r+1)}{k^2a^2J\,\xi} 
\;\,+ \frac{r^2}{3\,a^4H^2\Gamma}
\right)^{-1}\,.
\end{equation}
In the above, the first term dominates in the dRGT limit while the second term dominates in the large momentum limit. In order to control which limit is stronger, we define the quantity
\begin{equation}
\mathcal{E} \equiv \frac{k^2J}{a^2H^2}\,.
\end{equation}
The case $\mathcal{E}\ll 1$ then corresponds applying the dRGT limit first, while $\mathcal{E}\gg 1$ corresponds to applying the sub-horizon limit first. To make this argument clearer, we rewrite this new parameter in terms of the relevant length scales in the problem
\begin{equation}
\mathcal{E} = \frac{l_{H}\,l_{GMG}}{\lambda^2}\,,
\end{equation}
where $\lambda=a/k$ is the physical wavelength, the horizon length is $l_H\equiv 1/H$, while we define the length scale associated with varying mass parameters as $l_{GMG} \equiv J/H$. We summarise the possible values of the wavelength with respect to these scales in Figure~\ref{fig:scales}. For a small departure from constant mass dRGT, the two characteristic lengths obey $l_H >l_{GMG}$. For modes with wavelengths $\lambda \ll l_{GMG}$, the variation of the mass parameters is non-negligible, thus this case corresponds to the $\mathcal{E} \gg 1$ limit. For $l_{GMG} \ll \lambda \ll l_{H}$, the modes are sub-horizon, but the departure from standard dRGT is negligible, corresponding to $\mathcal{E} \ll 1$.

\subsubsection{Sound Speeds}
Instead of obtaining the full dispersion relations of eigenmodes, we will make use of the fact that the frequency is dominated by the gradient term at high momenta. We first vary the action (\ref{finalscalar}) with respect to the fields $\Psi^\dagger$, what results is the following equation of motion
\begin{equation}\label{eom3}
\mathcal{K}\,\ddot{\Psi}+\left(\frac{\mathcal{G}-\mathcal{G}^T}{2}+3H\mathcal{K}+\dot{\mathcal{K}}\right)\dot{\Psi}+\left(\frac{\dot{\mathcal{G}}}2+\frac{3\,H\,\mathcal{G}}{2}+\mathcal{M}\right)\Psi=0.
\end{equation}
For a monochromatic wave of the form $\Psi\propto e^{-i \int \omega dt}$, and in the sub-horizon limit we have $|\dot{\omega}| \ll\omega^2$, so  equation (\ref{eom3}) can be converted into an eigenvalue equation to solve $\omega$
\begin{equation}
\det\left[-\omega^2 \mathcal{K}-i\omega \left(\frac{\mathcal{G}-\mathcal{G}^T}{2}+3H\mathcal{K}+\dot{\mathcal{K}}\right)+\left(\frac{\dot{\mathcal{G}}}{2}+\frac{3H\mathcal{G}}{2}+\mathcal{M}\right)\right]=0\,.
\end{equation}
Since deep in the horizon, frequency is $\omega =C_S(k/a)$, we solve the above equation for the squared sound speed $C_S^2$. The equation is quadratic in $C_S^2$. The first solution coincides with the sound speed of the k-essence field
\begin{equation}
C_{S,1}^2=c^2_s.
\end{equation}
The second one provides the sound speed of the scalar graviton $C_{S,2}^2$, which is presented in Appendix \ref{append}. Both of  the squared-sound speeds should be positive to avoid gradient instability.

\section{Minimal Generalised Massive Gravity}\label{mini}
In this section we consider small departures from constant mass parameters by allowing only one of the $\alpha_n$ parameters to vary slowly. This allows us to solve the St\"uckelberg constraint \eqref{stuck} and evaluate the stability conditions in a concrete framework.
In this minimal setup, the free $\alpha_n(\phi^a\phi_a)$ functions in the action \eqref{model1} are
\begin{align}\label{lim}
\alpha_0(\phi^a\phi_a)&=\alpha_1(\phi^a\phi_a)=0\,,\nonumber\\
\alpha_2(\phi^a\phi_a)&=1+m^2\alpha_2' \phi_a\phi^a\,,\nonumber\\
\alpha_3(\phi^a\phi_a)&=\alpha_3 \,,\nonumber\\
\alpha_4(\phi^a\phi_a)&=\alpha_4 \,.\nonumber\\.
\end{align}
For the background configuration \eqref{fields}, we have $(\phi^a\phi_a) = -f(t)^2$. The above choice is basically the constant mass dRGT theory, with the only exception that we allowed $\alpha_2$ to vary with the St\"uckelberg fields. The variation is assumed to be small $\alpha_2'\ll 1$, so we expect the solutions to be close to dRGT. In this case, the contribution from mass term to the Friedmann equation $m^2 L$ should be approximately constant, and if it is responsible for late time acceleration, also positive. The conditions we impose are the positivity of the effective cosmological constant, the squared-tensor mass, vector and scalar gradient and kinetic terms.

We expand all background quantities for small $\alpha_2'$,
\begin{align}
\xi =& \xi_0 + \alpha_2' \xi_1 + \mathcal{O}\left(\alpha_2'\right)^2\,,\nonumber\\
J =& J_0 + \alpha_2' J_1 + \mathcal{O}\left(\alpha_2'\right)^2\,.\nonumber\\
\vdots
\end{align}
At leading order, i.e. $\mathcal{O}(\alpha_2')^0$, all background quantities reduce to standard (constant mass) dRGT expressions. The St\"uckelberg equation \eqref{stuck} at this order is simply $J_0= 0$, solved by \cite{Gumrukcuoglu:2011zh}
\begin{equation}
\xi_{0\,\pm} = \frac{1+2\,\alpha_3+\alpha_4 \pm \sqrt{1+\alpha_3+\alpha_3^2-\alpha_4}}{\alpha_3+\alpha_4}\,.
\label{eq:xpm}
\end{equation}
Moreover, since $\dot{\xi} = (-H + \sqrt{\kappa}\,r/a)\xi$, we can use that $\dot{\xi}_0=0$ to find
\begin{equation}
r_0 = \frac{a_0H_0}{\sqrt{\kappa}}\,.
\label{eq:r0}
\end{equation}
To determine which solution of $\xi_0$ is relevant, we can use  the squared tensor mass in the dRGT limit \cite{Gumrukcuoglu:2011zh},
\begin{equation}
\Gamma_0 = \pm \left(\frac{a_0H_0}{\sqrt{\kappa}}-1\right)\,\xi_{0\,\pm}^2\sqrt{1+\alpha_3+\alpha_3^2-\alpha_4}\,.
\end{equation}
Provided that curvature never dominates the expansion, the tensor mass is real only for the solution $\xi_{0\,+}$. We therefore consider this solution in the remainder of this section. 

We now discuss the parameter region where we have a real solution that leads to a positive cosmological constant. By definition, $\xi_0$ is a positive quantity, so the parameter region where the solution $\xi_{0\,+}$ in \eqref{eq:xpm} is positive corresponds to
\begin{equation}
\Big(\alpha_3 <-1 \land \alpha_4<-3(1+\alpha_3)\Big) \lor
\Big(\alpha_3 >-1 \land \alpha_4> -\alpha_3\Big) \,.
\end{equation}

On the other hand the contribution to the Friedmann equation from the mass term is then an effective cosmological constant
\begin{align}
L_0 &= (\xi_0-1)\left[6+4\,\alpha_3+\alpha_4 - (3+5\,\alpha_3+2\,\alpha_4)\xi_0 + (\alpha_3+\alpha_4)\xi_0^2\right]\,,
\nonumber\\
&= -\frac{1}{(\alpha_3+\alpha_4)^2}\left[(1+\alpha_3)(2+\alpha_3+2\,\alpha_3^2-3\,\alpha_4) + 2\,(1+\alpha_3+\alpha_3^2-\alpha_4)^{3/2}\right]\,.
\end{align}
In order to have a real and positive cosmological constant, we need to satisfy these conditions:
\begin{equation}
\alpha_3 >-1 \, \land \, \frac{3+2\,\alpha_3+3\,\alpha_3^2}{4}< \alpha_4 <1+\alpha_3+\alpha_3^2\,.
\label{eq:positivelambda}
\end{equation}

The $\mathcal{O}(\alpha_2')$ terms  are relevant only in the stability conditions, and they are exclusively introduced by the function $J$, whose $\mathcal{O}(\alpha_2')^0$ contribution vanishes. We can solve background equations \eqref{BGeqns} and \eqref{stuck} for $J_1$ to obtain
\begin{align}
J_1 =& \frac{2\,m^2\,a_0^2}{(\alpha_3+\alpha_4)^3\left(\sqrt{\kappa}a_0H_0-\kappa\right)}\Bigg[
(1+\alpha_3)\left(3\,\alpha_3^2+\alpha_3(5-2\,\alpha_4)-(\alpha_4-1)(\alpha_4+4)\right)
\nonumber\\
&\qquad\qquad\qquad\qquad\qquad\qquad
+\sqrt{1+\alpha_3+\alpha_3^2-\alpha_4}\,\left(4+\alpha_3(7+3\,\alpha_3)-\alpha_4-2\,\alpha_3\alpha_4-\alpha_4^2\right)
\Bigg]\,.
\end{align}
This is the main quantity needed when calculating the kinetic terms of perturbations. For the region \eqref{eq:positivelambda}, where cosmological constant and tensor mass is positive, the positivity of $J$ requires
\begin{equation}
\alpha_2' (\alpha_3-\alpha_4+1)>0\,.
\label{eq:Jpositive}
\end{equation}

We now discuss the conditions for avoiding ghost and gradient instability in the vector and scalar perturbations.
Schematically the action for the vector modes takes the form,
\begin{equation}\label{v1}
    S^{(2)}_V=\int d^3 k dt \,a^3\, \mathcal{T}_V\left(\vert\dot{V}\vert^2-
    \frac{c_{V}^2k^2}{a^2}\vert V \vert^2 + \dots\right)\,,
\end{equation}
where ellipsis denotes other terms in the action, e.g. mass. For the vector modes, both limits of the parameter $\mathcal{E}$ give the same result for the sub-horizon expressions, with
\begin{equation}
\mathcal{T}_V \to \frac{m^2 a_0^2\,J_1\,\xi_0\,\alpha_2'}{1+r_0}\,,\qquad
c_V^2 \to \frac{(1+r_0)\,\Gamma_0}{2\,J_1\,\xi_0\,\alpha_2'}\,.
\end{equation}
Since $r_0$ given by \eqref{eq:r0} is positive, and we chose the branch where $\Gamma_0>0$, avoiding both ghost and gradient instability requires
\begin{equation}
J_1\,\alpha_2' > 0\,,
\label{eq:vectorstability}
\end{equation}
which, in the regime where we have positive cosmological constant, corresponds to the range \eqref{eq:Jpositive}.

For the scalar mode that corresponds to the matter field, there is no ambiguity; as long as the equivalent fluid obeys the null energy condition, and has a real propagation speed, it is stable. For the scalar graviton, the action is formally
\begin{equation}\label{s1}
    S^{(2)}_S=\int d^3 k dt \,a^3\, \mathcal{T}_S\left(\vert\dot{S}\vert^2-
    \frac{c_{S}^2k^2}{a^2}\vert S \vert^2 + \dots\right)\,.
\end{equation}
As discussed in the previous section, the sub-horizon limit for $\mathcal{T}_S$ depends also on the limit for the parameter $\mathcal{E}$. We find
\begin{equation}
\mathcal{T}_S \to\left\{
\begin{array}{ll}
 \dfrac{3\,m^2a_0^4\,H_0^2\,\Gamma_0}{M_p^2r_0^2}&\,,\;\;\mathcal{E} \gg 1\\\\
 \dfrac{m^2a_0^2k^2\xi_0J_1\alpha_2'}{2\,M_p^2(r_0+1)}&\,,\;\;\mathcal{E} \ll 1\\
\end{array}
\right.\,.
\end{equation}
On the other hand, the sound speed for the scalar graviton has the same value when $\mathcal{E}$ is sent to the either of the two extremes,
\begin{equation}
c_S^2 \to \frac{2(1+r_0)\,\Gamma_0}{3\,J_1\,\xi_0\alpha_2'} = \frac{4}{3}\,c_V^2.
\end{equation}
Regardless of the $\mathcal{E}$ limit, scalar kinetic term and squared-sound speed are positive in the region where cosmological constant is positive, tensor mass is real ($\Gamma_0>0$) and the vector mode stability condition \eqref{eq:vectorstability} is satisfied.

\begin{figure}[ht]
    \begin{center} 
    ~\!\!\!\!\!\!\!\!\!\!\!\!\!\!\!\includegraphics[width=0.5\columnwidth]{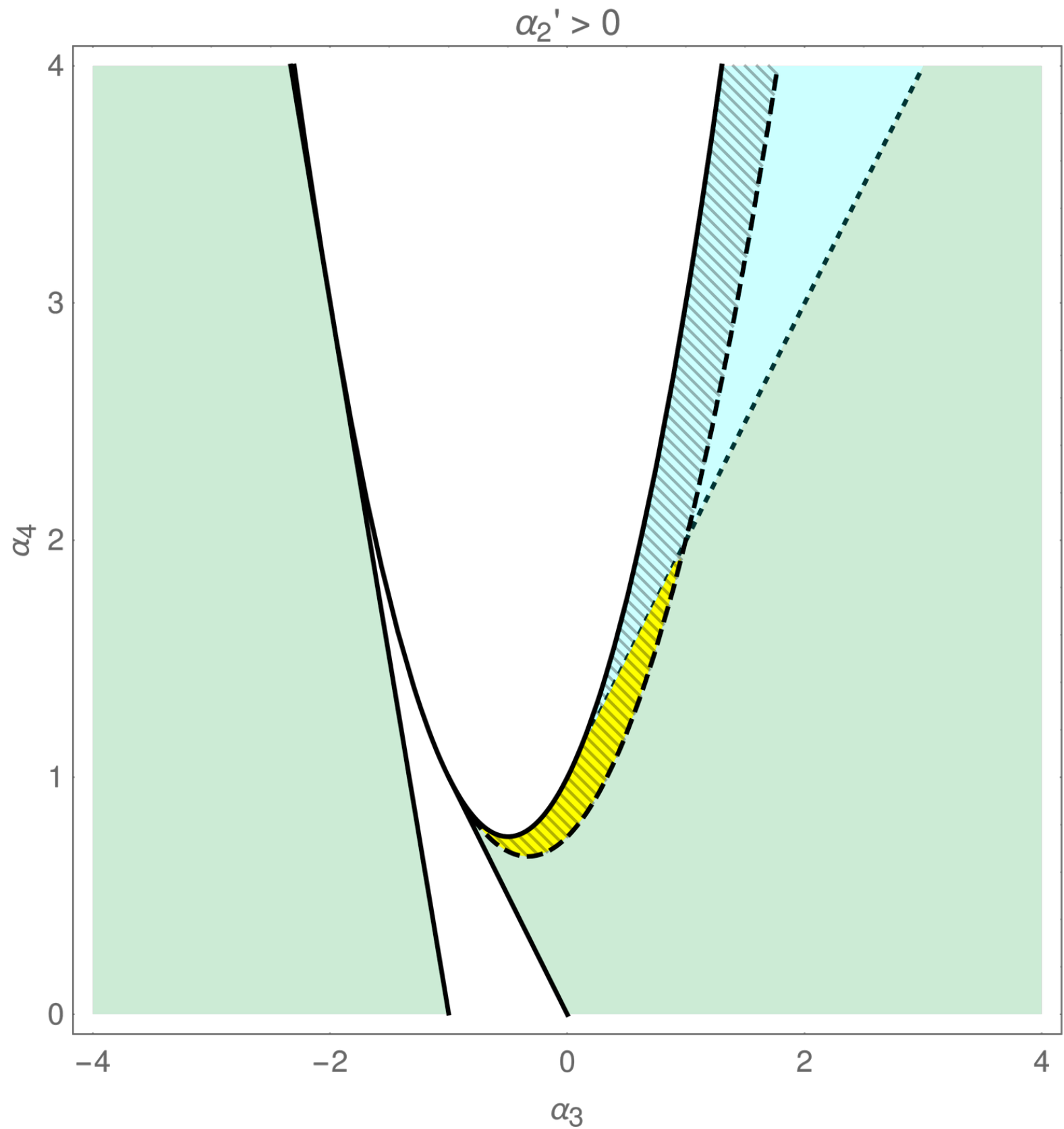}~~~~~~~~
    \includegraphics[width=0.5\columnwidth]{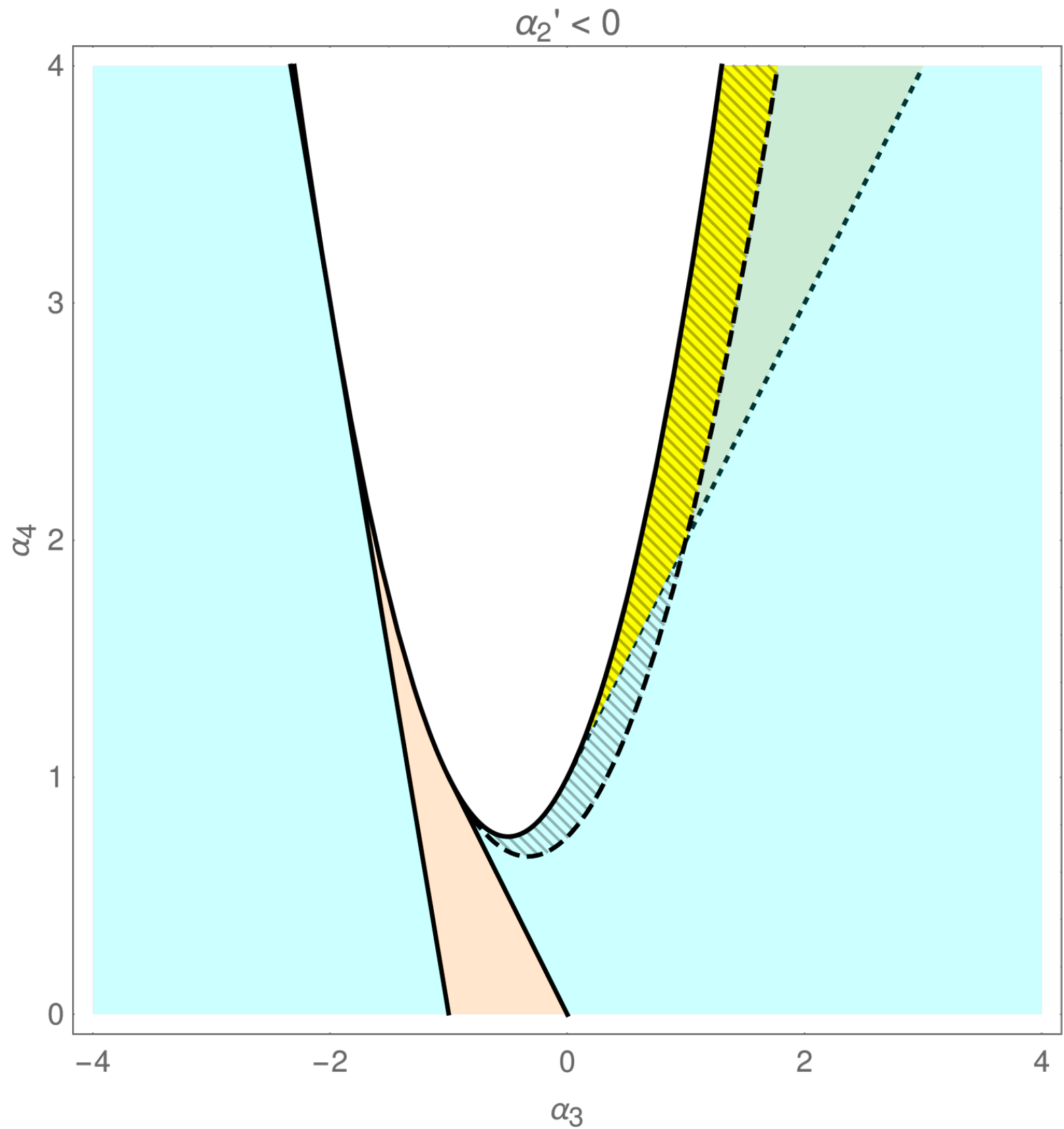}
    \end{center}
    \caption{Allowed regions for $\alpha_2' >0$ (left panel) and $\alpha_2' <0$ (right panel). The blue region is $\xi_0>0$ (bounded by the solid lines), orange region is $\alpha_2'J_1 > 0$ (bounded with the solid and the dotted lines), the green region is where both conditions are satisfied. We also mark the positive cosmological constant as the shaded area (region between the solid and dashed lines). The region where all conditions are satisfied is highlighted in yellow.}
    \label{fig:region}
\end{figure}

In Fig.\ref{fig:region}, we summarise all the conditions obtained in this section for the minimal model. 
We show the region of the parameter space that has positive cosmological constant and stable perturbations. Depending on the sign of $\alpha_2'$ parameter, the allowed region is either constrained to a finite area ($\alpha_2'>0$) or is open ($\alpha_2'<0$).

\section{Discussion and Conclusions}\label{conc}
In this paper we have studied the cosmological perturbations of Generalised Massive Gravity with a k-essence fluid as the matter sector. We calculated the quadratic action for the tensor, vector and scalar sectors and identified the stability conditions. We found that, unlike in constant mass dRGT massive gravity, the kinetic terms for the vector and scalar gravitons are non-vanishing, and the background can be free from pathologies. As an example, we introduced a minimal version of the theory where only one mass function is allowed to vary slowly, which can be considered as a small variation from standard dRGT. In this model, the contribution to the Friedmann equation from the mass term is approximately a constant that can be positive. In other words, the cosmology approximates GR with a cosmological constant. On the other hand, the effective cosmological constant continues to vary, and this variation allows the background to be perturbatively stable in a region of the parameter space, unlike constant mass dRGT theory. The tensor graviton has a time dependent mass and propagates at the speed of light, while vector and scalar perturbations generically propagate at superluminal speeds.

The cosmology of Generalised Massive Gravity has advantages over similar extensions. In addition to 
being an extension that has the same number of degrees of  freedom as standard dRGT, the strong coupling problem can be tamed, unlike some other extensions.
For instance, the problem of vanishing kinetic terms have been addressed in a similar manner in the mass-varying massive gravity \cite{Huang:2012pe}, where the mass parameters are promoted to functions of a new dynamical field. On the other hand, in order to achieve self-acceleration the mass functions need to vary slowly, making the scalar and vector modes strongly coupled \cite{Gumrukcuoglu:2013nza}. In contrast, in the present paper, we showed that the scalar perturbations in the GMG theory effectively end up with finite kinetic terms provided that one considers modes that are sub-horizon and below the characteristic GMG length scale. This is an indication that the strong coupling problem of the dRGT theory becomes milder even with a slight variation of the model parameters.

Finally, a study of perturbations in the GMG theory was performed in \cite{deRham:2014gla}. Although our results are qualitatively compatible, the quantitative connection is unclear. We think that the apparent discrepancy is due to their choice of Fermi normal coordinates and the decoupling limit potentially probing a different background than the one considered here.

As we have established a simple model of a stable cosmology with (approximate) self-acceleration, the next step would be to determine whether a sensible expansion history can be consistently accommodated in this framework. Another path would be to determine the details of Vainshtein screening in a study of the non-linear perturbations.

\label{BG}

\acknowledgments
We thank Obinna Umeh for helpful comments and discussions with respect to the tensor package xPand \cite{Pitrou:2013hga} which was used to perturb tensorial expressions in the calculations. We also acknowledge xTras \cite{Nutma:2013zea}. We thank Matteo Fasiello for englightening discussions. MK-A also thanks Chris Pattison and Mike Wang for helpful discussions. The work of AEG and KK has received funding from the European Research Council
(ERC) under the European Union’s Horizon 2020 research and innovation programme (grant agreement No. 646702 ”CosTesGrav”). KK is supported by the UK STFC ST/S000550/1.

\appendix
\section{Kinetic and gradient terms for the scalar sector}
\label{append}
In the scalar sector, the two eigenvalues are given by
\begin{align}
e_1 =& \left[\frac{M_p^4 c_s^2 H^2}{2\,(\rho+P)}
-\left(
\frac{4\,\left[2\,(k^2+3\,\kappa)\,(r+1)+3\,m^2a^2J\,\xi\right]}{M_p^2\left[2\,\kappa\,(r+1)+m^2\,a^2J\,\xi\right]}
+\mathcal{B}_1\right)^{-1}
\right]^{-1}\,,\nonumber\\
e_2 = & \Bigg\{
\frac{3\,M_p^2}{k^2(k^2+3\,\kappa)} + \frac{2\,M_p^2(r+1)}{k^2m^2a^2J\,\xi} \nonumber\\
&\;\,+ \frac{4\,M_p^2r^2}{3\,m^2a^2}
\left[
J\,\xi\left[-2\,\sqrt{\kappa}\,a\,\left(H-\frac{\sqrt{\kappa}}{a}\right)r+a^2\left(-4\,H^2+r\left(\frac{\rho+P}{M_p^2}+m^2J\,\xi\right)\right)\right]+2\,a^2H(2\,H\,\Gamma-\dot{J}\,\xi)
\right]^{-1}
\Bigg\}^{-1}\,,
\label{eq:kinetic-exact}
\end{align}
where
\begin{align}
\mathcal{B}_1 \equiv& \frac{4\,m^2J^2r^2(r+1)\,\xi^2 \left[6\,\kappa+2\,(k^2+3\,\kappa)r+3\,m^2a^2J\,\xi\right]^2}{M_p^2\left[2\,\kappa\,(r+1)+m^2a^2J\,\xi\right]}\nonumber\\
&\times \Bigg\{m^2J^2r\,\xi^2\left[-2\,k^2r^3-6\,\kappa(r-1)(1+r)^2 -3\,m^2a^2J(r^2-1)\,\xi \right]\nonumber\\
&\qquad+H\,\xi\left[2\,\kappa\,(r+1)+m^2a^2J\,\xi\right]\left[\mathcal{B}_2 J -6\,(r+1)\left(-\frac{2\,H\,\Gamma}{\xi}+\dot{J}\right)\right]\Bigg\}^{-1}\,,\nonumber\\
\mathcal{B}_2 \equiv & 
-\frac{3\,r\,(r+1)}{H^2}\left[\frac{2\,\sqrt{\kappa}\,H}{a}+\frac{4\,H^2}{r} -\frac{2\,\left[3\,\kappa+(k^2+3\,\kappa)r\right]}{3\,a^2(1+r)} - \frac{\rho+P}{M_p^2}
\right]\,.
\end{align}

The sound speed of the scalar graviton mode is given by 
\begin{align}
C_S^2 =&\frac{2}{3}\left(1-\frac{\Gamma}{J\,\xi}\right) + \frac{2}{3\,\mathcal{B}_3}\Bigg\{r\,\left(\xi\,\ddot{J}-2\,H\,\dot{\Gamma}\right)-\frac{(r+2)}{J\,\xi}\,\left( \xi\,\dot{J}-2\,H\,\Gamma\right)^2+\frac{\dot{J}}{J}\left(\xi\,\dot{J}-2\,H\,\Gamma\right)
\nonumber\\
&\qquad\qquad\qquad\qquad\qquad
-\Gamma\,\left[2\,H^2(r-8)-\frac{2\,\sqrt{\kappa}\,r}{a}\left((2\,r+3)H-\frac{\sqrt{\kappa}}{a}\,r\right)+3\,m^2J\,r^2\,\xi-2\,H\,\dot{r}\right]
\nonumber\\
&\qquad\qquad\qquad\qquad\qquad
+\frac{J\,\xi}{2}\left[
\frac{2\,\sqrt{\kappa}H\,r\,(r-6)}{a}-\frac{2\,\kappa\,r^2}{a^2}
+8\,H^2(r-2)+3\,m^2J\,r^2\xi-4\,H\,\dot{r}\right]
\nonumber\\
&\qquad\qquad\qquad\qquad\qquad
+\dot{J}\,\xi\left[2\,(r-3)\,H-\frac{2\,\sqrt{\kappa}\,r}{a}-\dot{r}\right]
\Bigg\}\,,
\end{align}
where
\begin{align}
\mathcal{B}_3 \equiv 2\,H\,(\xi\,\dot{J}-2\,H\,\Gamma)+2\,H\,J\,\xi\,\left(2\,H+\frac{\sqrt{\kappa}\,r}{a}\right)-J\,r\,\xi\,(m^2J\,\xi-2\,\dot{H})\,.
\end{align}

\bibliography{refGM}

\end{document}